\documentclass[fleqn,twoside]{article}
\usepackage{espcrc2}

\newcommand{\sbar}{\overline{s}}
\newcommand{\qbar}{\overline{q}}
\newcommand{\dbar}{\overline{d}}
\newcommand{\psibar}{\overline{\psi}}
\newcommand{\qt}{\tilde{q}}
\newcommand{\dt}{\tilde{d}}
\newcommand{\dtbar}{\overline{\dt}}
\newcommand{\str}{{\rm str}\,}
\newcommand{\betahat}{\hat{\beta}}

\newcommand{\AmS}{{\protect\the\textfont2
  A\kern-.1667em\lower.5ex\hbox{M}\kern-.125emS}}

\title{On systematic errors due to quenching in $\varepsilon'/\varepsilon$}

\author{Maarten Golterman\address[SFSU]{Department of Physics and Astronomy,
San Francisco State University, San Francisco, CA 94132, USA} and
        Elisabetta Pallante\address[SISSA]{SISSA, Via Beirut 2-4, I-34013
	Trieste, Italy}}

\begin{document}

\begin{abstract}
Recently, we pointed out that chiral transformation properties of penguin
operators change in the transition from unquenched to (partially) quenched
QCD. As a consequence, new operators appear in (partially) quenched
QCD penguins, introducing ambiguities which should be considered a quenching
artifact.  Here we discuss more specifically the
effects of this phenomenon on the quenched $\Delta I=1/2$ $K\to\pi\pi$ amplitude,
and in particular, its potential numerical effect on recent lattice estimates 
for $\varepsilon'/\varepsilon$.
\vspace{1pc}
\end{abstract}

\maketitle

The strong $LR$ penguin operators $Q_{5,6}$ mediating $\Delta S=1$,
$\Delta I=1/2$ weak transitions are of the form 
\begin{equation}
Q^{QCD}_{penguin}\propto(\sbar\gamma_\mu(1-\gamma_5)d)(\qbar
\gamma_\mu(1+\gamma_5)q)\ ,
\end{equation}
with $q$ summed over $u$, $d$, $s$, and thus transform trivially
under SU(3)$_R$.  However, this is no longer true when one quenches
the theory.  The quenched theory can be defined by introducing a
bosonic ``ghost" quark $\qt$ for each of the three valence quarks $q$ \cite{morel},
and thus the symmetry group is enlarged to a graded group
transforming all six quarks into each other, SU(3)$_L\times$SU(3)$_R\to$
SU(3$|$3)$_L\times$SU(3$|$3)$_R$ \cite{bg}, under which $Q^{QCD}_{penguin}$ is
no longer a singlet.  This was observed in ref.~\cite{gp}, in which also
the consequences for $K\to 0$ and $K\to\pi$ matrix elements were studied
in ChPT, both in the fully and partially quenched cases.  Here we extend
the discussion to $K\to\pi\pi$ amplitudes, restricting ourselves to the
quenched three-flavor theory.  The partially quenched case, as well as more
general results, are contained in ref.~\cite{gp2}.

The right-handed current in eq.~(1) can be split into a singlet ($S$) and a
non-singlet ($NS$) part, $(\qbar q)_R\to\frac{1}{2}(\psibar\psi)_R
+\frac{1}{2}(\psibar{\hat N}\psi)_R$ (in an obvious shorthand), 
where $\psi$ runs over valence {\it and}
ghost quarks, and ${\hat N}={\rm diag}(1,1,1,-1,-1,-1)$.  Accordingly,
$Q^{QCD}_{penguin}$ splits up into a singlet and a non-singlet part
$Q^{QS}$, resp. $Q^{QNS}$, which are represented
in leading-order ChPT by
\begin{eqnarray}
&&\hspace{-0.7cm}Q^{QS}\to-\alpha^{(8,1)}_{q1}\str(\Lambda L_\mu L_\mu)
+\alpha^{(8,1)}_{q2}\str(\Lambda X_+)\ ,\nonumber \\
&&\hspace{-0.7cm}Q^{QNS}\to
f^2\alpha^{NS}\str(\Lambda\Sigma{\hat N}\Sigma^\dagger)\ ,
\end{eqnarray}
with $\alpha^{NS}$ a new low-energy constant (LEC), which only exists
in the quenched theory, and thus should be considered an artifact of
the quenched approximation \cite{gp}.\footnote{See also this reference for 
notation.}  While a similar phenomenon also occurs for $LL$ penguins,
we note that in the $LR$ case, the new non-singlet operator is $O(p^0)$
in ChPT, thus enhancing its contribution to weak matrix elements.  (This
is similar to the enhancement of the EM penguins $Q_{7,8}$.)  

At tree level, the new operator does not contribute to matrix elements
with only physical (valence) mesons on the external lines, but it does
contribute at one loop, thus competing at $O(p^2)$ with the 
{\it tree-level} contributions from $Q^{QS}$.  For the $K\to\pi$ (with
degenerate masses) and
$K\to 0$ matrix elements used recently by the CP-PACS \cite{cppacs}
and RBC \cite{rbc} collaborations to extract the value of $\alpha^{(8,1)}_{q1}$
we find \cite{gp}
\begin{eqnarray}
&&\hspace{-0.7cm}\langle\pi^+|Q^{QCD}_{penguin}|K^+\rangle= \\
&&\hspace{-0.7cm}\frac{4M^2}{f^2}\left[(\alpha^{(8,1)}_{q1}
\!-\betahat^{NS}_{q1}\!-\frac{1}{2}\betahat^{NS}_{q2})-
(\alpha^{(8,1)}_{q2}+\betahat^{NS}_{q3})\right],\nonumber\\
&&\hspace{-0.7cm}\langle 0|Q^{QCD}_{penguin}|K^0\rangle= \\
&&\hspace{-0.7cm}\frac{4i}{f}\left[(M_K^2-M_\pi^2)
(\alpha^{(8,1)}_{q2}+\betahat^{NS}_{q3})
+\alpha^{NS}*{\rm logs}\right]\ ,\nonumber
\end{eqnarray}
and for the $K\to\pi\pi$ matrix element with physical kinematics
we find \cite{gp2}
\begin{eqnarray}
&&\hspace{-0.7cm}\langle\pi^+\pi^-|Q^{QCD}_{penguin}|K^0\rangle=
\frac{4i}{f^3}\Bigl[\\
&&\hspace{-0.7cm}(M_K^2-M_\pi^2)(\alpha^{(8,1)}_{q1}
\!-\betahat^{NS}_{q1}\!-\frac{1}{2}\betahat^{NS}_{q2})
\!+\alpha^{NS}\!*{\rm logs}\Bigr].\nonumber
\end{eqnarray}
Here $\betahat^{NS}_{qi}\equiv(4\pi)^2\beta^{NS}_{qi}$ are $O(p^2)$ LECs
which appear at next-to-leading order in the non-singlet sector, and absorb 
the scale dependence from the chiral logs \cite{gp}.  
Note that the singlet and non-singlet
LECs always appear in the same linear combinations.  This can be
understood from the fact that only physical mesons appear on the
external lines (see ref.~\cite{gp2} for details).

{}From these results, we conclude that there are at least three
different strategies one could follow to estimate the real-world 
$\Delta I=1/2$ decay rate from a quenched computation:\\ \\
$\bullet$ Ignore $\alpha^{NS}$, but not the $\betahat^{NS}_{qi}$.  
This introduces a scale dependence,
but makes sense if $\alpha^{NS}\!*{\rm logs}$ is small
in the physical matrix element, at a reasonable scale.
Thus $\alpha^{(8,1)}_{q1}-\betahat^{NS}_{q1}-\frac{1}{2}\betahat^{NS}_{q2}$
from $K\to\pi$ and $K\to 0$
is taken as the ``best" estimate of unquenched $\alpha^{(8,1)}_1$,
and the physical $K\to\pi\pi$ matrix element is calculated from the
unquenched expression $4i(M_K^2-M_\pi^2)\alpha^{(8,1)}_1/f^3$.
This is the strategy followed by CP-PACS and RBC \cite{cppacs,rbc}.\\
$\bullet$ Drop all the non-singlet operators, following the simple prescription
given in ref.~\cite{gp}.  This was investigated for $Q_6$ in ref.~\cite{lanl};
see below. \\
$\bullet$ Keep everything in eqs.~(3-4), and also calculate the physical 
$K\to\pi\pi$ matrix element using eq.~(5), assuming that it gives the best 
estimate of the real-world matrix element. \\

Naively, the second strategy might be considered the obvious one, 
since the aim is to compute the (unquenched) value of $\alpha^{(8,1)}_1$,
which (at least at tree level in unquenched ChPT) determines the penguin
contribution to the physical decay rate.  However, we do not know whether
the quenched (e.g. $\alpha^{(8,1)}_{q1}$) and unquenched 
(e.g. $\alpha^{(8,1)}_1$) LECs have the same values; it might happen that 
$\alpha^{(8,1)}_{q1}-\betahat^{NS}_{q1}-\frac{1}{2}\betahat^{NS}_{q2}$ 
is a better estimate of $\alpha^{(8,1)}_1$.  For this reason, we believe that 
the
spread in values obtained by following different strategies should be taken as
a {\em systematic error} due to quenching.  
The origin of this uncertainty comes
from the fact that in the case of electro-weak operators, defining the
quenched theory as QCD with the fermion determinant set equal to a constant
is {\it not} enough: in addition, a prescription has to be given for the
``embedding" of electro-weak operators in the quenched theory \cite{gp,ms}.
In some cases this is straightforward, but in the case at hand it
is not.  Note that in the case of partially quenched QCD with three dynamical
light flavors, the choice of strategy {\it is} unambiguous: the second 
strategy should be followed \cite{gp}, 
since the LECs of this theory are those of unquenched QCD \cite{ss}.

We will now comment on the three different strategies listed above.
First, we address the issue of ignoring $\alpha^{NS}$.  We found that
if $\alpha^{NS}$, with the normalization defined in eq.~(2), is of the
same order as $\alpha^{(8,1)}_{q1}$, its numerical contribution to the
$K\to\pi\pi$ matrix element in eq.~(5) is small ($\sim 10^{-2}$ of the total),
for a reasonable choice
of the ChPT scale~\cite{gp2}.  If so, this justifies ignoring $\alpha^{NS}$.
The reason appears to be purely numerical, and can be understood as a 
suppression by the typical factor $1/(4\pi)^2$ arising at one loop.

It would of course be better to determine $\alpha^{NS}$ from a lattice
computation.  In principle, it can be determined from existing data
for the $K\to 0$ matrix element ({\it cf.} eq.~(4)), but this may be
difficult to do in practice, since it is hard to disentangle logs from
the linear term in the quark mass at the typical values of quark masses 
employed
in current simulations \cite{cppacs,rbc}. In fact, it is easier to 
determine $\alpha^{NS}$ from a process in which it appears at leading
order in ChPT.  This can be done as follows.  First, rotate 
$Q^{QCD}_{penguin}$ by a SU(3$|$3)$_L$ rotation into
$(\sbar\gamma_\mu(1-\gamma_5)\dt)(\psibar{\hat N}\gamma_\mu
(1+\gamma_5)\psi)$, which is in the same representation of
SU(3$|$3)$_L\times$SU(3$|$3)$_R$.  Next, consider the matrix element
between a fermionic kaon ${\tilde K}\propto\dtbar\gamma_5 s$
and the vacuum.  ChPT tells us that
\begin{equation}
\langle 0|(\sbar\dt)_L(\psibar{\hat N}\psi)_R|{\tilde K}\rangle
=2if^2\alpha^{NS}+O(p^2).
\end{equation}
The key observation is that no simulations with ghost quarks are
needed.  After carrying out all Wick contractions on the left-hand side
of eq.~(6), one simply uses the fact that ghost and valence propagators
are equal, $\langle\dt(x)\dtbar(y)\rangle=\langle d(x)\dbar(y)\rangle$,
{\it etc}.  Estimating $\alpha^{NS}$ is thus as simple as estimating
$\alpha^{(8,1)}_{2q}$, once the usual valence-quark propagators have been
computed.  It is important to determine $\alpha^{NS}$ in order to see whether
indeed its contribution can be considered a small effect.

The remaining choice of strategy boils down to whether one considers
$\alpha^{(8,1)}_{1q}$ or $\alpha^{(8,1)}_{1q}\!-\!\betahat^{NS}_{q1}
\!-\!\frac{1}{2}\betahat^{NS}_{q2}$ to be a better estimate of unquenched
$\alpha^{(8,1)}_1$.  Since the issue only arises for QCD penguins,
and is a non-leading effect in ChPT for the $LL$ case \cite{gp},
it is unlikely to have much influence on the
real part of the $\Delta I=1/2$ amplitude.  This is not the
case for $\varepsilon'/\varepsilon$, where $Q_6$ is expected to be one of the
largest contributions.  In ref.~\cite{lanl} it was found,
following the simple prescription for dropping non-singlet operators
given in ref.~\cite{gp} (and using staggered
fermions, while refs.~\cite{cppacs,rbc} used domain-wall fermions),
that leaving out the non-singlet operator in $Q_6$ enhances $B_6^{(1/2)}$
by a factor two at the kaon mass.  To leading order in ChPT, this means
that $\alpha^{(8,1)}_{1q}$ is twice as large as 
$\alpha^{(8,1)}_{1q}\!-\!\betahat^{NS}_{q1}\!-\!\frac{1}{2}\betahat^{NS}_{q2}$,
potentially enhancing the contribution of $Q_6$ to $\varepsilon'/\varepsilon$ 
by a factor two relative to the contributions found in
refs.~\cite{cppacs,rbc}.

No estimate of this enhancement with domain-wall fermions
is available.  In order to get an idea of how large the effect on
$\varepsilon'/\varepsilon$ can be, we therefore consider what a factor-two
enhancement of $B_6^{(1/2)}$ would imply for the central value of
$\varepsilon'/\varepsilon=-4\times 10^{-4}$ reported by RBC \cite{rbc}.
We find that this value would shift to $+10\times 10^{-4}$.
Considering the CP-PACS results \cite{cppacs} the situation is less
clear.  They typically find a smaller (in absolute value) negative
contribution from $Q_6$, and there appears to be a stronger dependence
on quark mass.  When we take their value of $-2\times 10^{-4}$ at
a degenerate Goldstone-boson mass of about 600~MeV, we find that enhancing
$B_6^{(1/2)}$ by a factor two changes this into $+2\times 10^{-4}$.

We emphasize that these are rather unreliable estimates of the effect
of keeping or dropping the non-singlet operators in $Q_6$.  Many
other effects are not under control, such as the fact that all
computations were done at only one value of the lattice spacing,
and that our estimates rely on leading-order ChPT, to mention but a few.
However, these estimates do make it clear that the issue is important,
not only theoretically, but also phenomenologically.  While the
quenched approximation does remarkably well in ``simple" strong
interaction physics, such as the light hadron spectrum, it is
a major obstacle to obtaining phenomenologically relevant estimates
of finely-tuned quantities like $\varepsilon'/\varepsilon$.  
Partially quenched computations with three dynamical light flavors are needed
in order to resolve the ambiguities introduced by quenching.

Finally, we recall that similar ambiguities also affect $Q_5$ and
strong $LL$ penguins.  However, these are expected to be numerically
less important.  For $Q_5$ the effect occurs at leading order in ChPT,
just as for $Q_6$, but the corresponding Wilson coefficient in the
weak $\Delta S=1$ hamiltonian is much smaller, suppressing the
contribution of $Q_5$ altogether.  For $LL$ penguin operators, the
effect is non-leading in ChPT, and affects $K\to\pi\pi$ matrix elements
only at $O(p^4)$. 

{\it Acknowledgements.} We would like to thank N. Christ, R. Mawhinney,
M. Savage and S. Sharpe for discussions.  EP thanks the Dept. of Physics and
Astronomy
of San Francisco State University for hospitality.  MG is supported in part by
the US Dept. of Energy; EP was supported in part by the Italian MURST under 
the program \textit{Fenomenologia delle Interazioni Fondamentali}.

\end{document}